\documentclass[aps,twocolumn,groupedaddress,showpacs,prb,floatfix]{revtex4}
\usepackage{graphicx,rotating,subfigure,amsmath,amsfonts,amssymb,delarray} 
\usepackage{color}
\usepackage{graphicx}
\usepackage{dcolumn}
\usepackage{bm}
\usepackage{epsfig}
\usepackage{umlaut}

\newcommand{\tbl}[1]{\textcolor{blue}{#1}}
\renewcommand{\vec}[1]{\boldsymbol #1}
\newcommand{\be}{\begin{eqnarray}}
\newcommand{\ee}{\end{eqnarray}}

\def\refeq#1{(\ref{#1})}
\def\d{\mbox d}

\def\nn{\nonumber}

\def\L{\Lambda}

\def\Or{\mathcal O}

\def\al{\alpha}

\def\o{\omega}

\def\ve{\varepsilon}
\def\l{\left}
\def\r{\right}
\def\te{\mbox{e}}
\def\rmi{{\rm i}}

\def\up{\uparrow}
\def\Up{\Uparrow}
\def\down{\downarrow}
\def\Down{\Downarrow}

\def\det{\mbox{det}}
\def\M{\mathcal{M}}

\begin{document}
\bibliographystyle{apsrev}
\title{Exact dynamics in the inhomogeneous central-spin model}
\author{Michael Bortz}
\email{michael.bortz@anu.edu.au}
\affiliation{Department of Theoretical Physics, Research School of Physics and Engineering, Australian National University, Canberra ACT 0200, Australia}
\author{Joachim Stolze}
\affiliation{Institut f\"ur Physik, Universit\"at Dortmund, 44221 Dortmund, Germany}
\date{\today}
\begin{abstract}
We study the dynamics of a single spin-1/2 coupled to a bath of spins-1/2 by inhomogeneous Heisenberg couplings {including} a central magnetic field. This central-spin model describes decoherence in quantum bit systems. An exact formula for the dynamics of the central spin is presented, based on the Bethe ansatz. For initially completely polarized bath spins and small magnetic field we find persistent oscillations of the central spin about a nonzero mean value.
For a large number of bath spins $N_b$, the oscillation frequency is proportional to $N_b$, whereas the amplitude behaves like $1/N_b$, to leading order. No asymptotic decay of the oscillations due to the non-uniform couplings is observed, in contrast to some recent studies. 
\end{abstract}
\maketitle
\section{Introduction}
The central-spin model, or Gaudin model {\cite{gau76,gaubook}}, is defined through the Hamiltonian 
\be
H=\sum_{j=1}^{N_b}A_j \vec S_0 \cdot \vec S_j - h S_0^z\label{gaudef}\,.
\ee
One central spin $\vec S_0$ interacts with $N_b$ bath spins via inhomogeneous Heisenberg couplings $A_j$. Additionally, a magnetic field $h$ is included that couples to the central spin only. Here, we restrict ourselves to spin-1/2 objects. This model is being used widely to model the hyperfine interaction between an electron spin in a quantum dot and the surrounding nuclear spins \cite{sch02,sch03,dob03,sem03,kha02,kha03,den06,has06,coi06}. The 
{localization of the electron wave function within}
the quantum dot leads to inhomogeneous coupling constants, depending on the distance of the nuclei from the center of the dot \cite{sch03,kha03}. Since {electrons in} quantum dots are candidates for spin qubits \cite{los98}, special interest is on decoherence phenomena of the central spin. In \cite{kha02,kha03} it was argued that the non-uniform hyperfine couplings cause the central spin to decay. Two special cases were considered in those works: a completely polarized and a completely unpolarized bath. In the case of a strong $h$-field, results for arbitrary polarization have been obtained in \cite{coi04}. {In all cases, the initial oscillations of the central spin were reported to decay, implying that the central spin comes to rest at some equilibrium value} due to the inhomogeneous couplings to the environmental spins. In this paper, we show that at least for an initially completely polarized bath, {the exact solution from the Bethe ansatz (BA) behaves differently}. 

The model Eq.~\refeq{gaudef} has also long been of interest from a fundamental point of view. As mentioned above, its eigenvalues and eigenstates can be constructed exactly in the framework of the BA \cite{gau76,gaubook,skl89,gau95}. In this formalism, the solution of the many-particle eigenvalue problem is achieved by
the solution of coupled algebraic equations for the quantum numbers (BA numbers). Both the eigenstates and the eigenvalues are constructed in terms of these numbers. {However, no BA studies of the dynamics have been performed yet, which is mainly due to the {intricate} distribution of the BA numbers in the complex plane.}

{Besides the full quantum-mechanical solution, a complementary approach consists in treating the model \refeq{gaudef} on a mean-field level, where operators are replaced by their expectation values. It was argued \cite{coi06,yuz05b} that this approach is exact in the thermodynamic limit. But, as pointed out very recently \cite{bor06}, agreement between the quantum and the mean-field solutions depends on the choice of the initial conditions in the quantum-mechanical solution. In \cite{bor06} the question was {addressed} to what extent initial entanglement between the bath spins is {essential for} the mean-field behaviour. If all couplings are equal, $A_j=:A\,\forall j$ in \refeq{gaudef}, then in the extreme case of an initially unentangled unpolarized bath the central spin was observed to decay from its initial {value} $\pm 1/2$ to a value $\sim 1/N_b$ after a decoherence time $\tau_d\sim 1/\sqrt{N_b}$. In other words, in the thermodynamic limit, {decoherence occurs infinitely
  fast} and after that, the dynamics is frozen. The aim of {the present} paper is to shed more light on the time evolution starting from an unentangled bath state for the physically relevant case of inhomogeneous couplings, using the exact BA solution.}

During the past few years, a whole variety of methods has been applied to study decoherence phenomena in the central-spin model. However, the potential of the BA method is still unexplored. {In this regard,} this work constitutes a first step to demonstrate the practicability of the BA approach by first presenting a general formula for the dynamics of the central spin and then focusing on one special case. The study of this special case should be considered not as the end, but rather as the starting point of the investigation of the exact solution. 

In the following, we derive an exact formula for the time-dependent expectation value $\langle S_0^z\rangle(t)$ of the central spin by employing the BA solution. {That formula is our central result. It is valid for an initial state which is a mutual eigenstate of all $S_j^z$ operators and hence} a product state containing no entanglement at all. The degree of spin polarization of the initial bath state is an essential control parameter.
For $M_b$ flipped spins (as compared to the ferromagnetic ``all up'' state) in the bath and with the central spin flipped as well, the result {for $\langle S_0^z \rangle(t)$} is {expressed in terms of} a sum of harmonic oscillations with frequencies determined by the sum of {$M_b+1$} BA numbers, and with amplitudes given through the norms of the eigenfunctions, see Eqs. (\ref{aljb},\ref{sz}) below.  We evaluate 
{$\langle S_0^z \rangle(t)$ analytically for $M_b=0$ (fully polarized bath), with couplings distributed in an interval $\l]0,\mathcal A\r]$ ($\mathcal A =$const). Under these conditions,
$\langle S_0^z\rangle(t)$ oscillates with a frequency $\sim \frac12\sum_{j=1}^{N_b} A_j$ and an amplitude of order $\Or(1/N_b)$ about a mean value   $\overline{\langle S_0^z \rangle(t)} = -\frac12 + \Or(1/N_b)$. The amplitude initially decreases but stays constant after  a time $t_a\sim 4 \pi/\mathcal A$. This differs from the results of  \cite{kha02,kha03},  where the oscillations of $\langle S_0^z \rangle(t)$ were found to decay completely. Our analytical  evaluation of $\langle S_0^z \rangle(t)$ is confirmed by a numerical evaluation for zero field and $N_b=30$ bath spins.}

{An} analytical evaluation is also possible for {a} large magnetic
field ($|h| \gg N_b$), with similar results, namely, persistent oscillations
of amplitude $\Or(N_b/h^2)$. From the magnetic field dependence of the results
we conclude that there exists a resonant value for $h$, where $\langle
S_0^z\rangle(t)$ oscillates at the maximum amplitude {possible}. {This result is also
  confirmed by numerical evaluation for $N_b=30$.}

The remainder of this paper is organized as follows. In the next section, we sketch the BA setting and give the general formula for $\langle S_0^z\rangle(t)$. The third section contains both analytical and numerical results for a fully polarized bath. 

\section{The Bethe ansatz solution}
The Bethe ansatz solution of the model \refeq{gaudef} was found by Gaudin \cite{gau76,gaubook,gau95}. Before summarizing it, let us introduce the abbreviations $N_b \bar A :=\sum_{j=1}^{N_b}A_j$, $N:=N_b+1$, {$M:=M_b+1$}, $C^N_M:=N!/(M!(N-M)!)$. The eigenvalues $\L_\nu$ of the Hamiltonian \refeq{gaudef} for a fixed number $M$ of down-spins (i.e. for magnetization $N/2-M$) were given in \cite{gaubook,skl89} in terms of the BA numbers $\o_{k,\nu}$: \footnote{Note that in those works, the inverse numbers $\ve_j:=1/A_j$, $E_{k,\nu}:=1/\o_{k,\nu}$ are employed.}
\be
\L_\nu&=&-\frac12 \sum_{k=1}^{M}\omega_{k,\nu}+\frac{N_b}{4}\bar A-\frac{h}{2}\,;\;\; \nu=1,\ldots,C^N_M\;,\label{ev}
\ee
provided that the $\omega_{k,\nu}$ fulfill the equations
\be
1+\sum_{j=1}^{N_b}\frac{A_j}{A_j-\omega_{k,\nu}}-2 \sum_{k'\neq k}^{M}\frac{\o_{k',\nu}}{\o_{k',\nu}-\o_{k,\nu}}+\frac{2 h}{\o_{k,\nu}}=0\nn 
\ee
for $k=1,\ldots,M$. Gaudin \cite{gau95} showed that there are $C^N_M$ sets of solutions $\l\{\o_{1,\nu},\ldots,\o_{M,\nu}\r\}$ to these equations, one for each eigenvalue $\Lambda_{\nu}$. The corresponding energy eigenstates with a fixed number $M$ of flipped spins read
\be
 |M_\nu\rangle&=&\frac{1}{n_{M_\nu}}\prod_{k=1}^{M}\sum_{j=0}^{N_b}\frac{A_j}{\o_{k,\nu}-A_{j}}S_j^-|0\rangle\nn\\
&=& \frac{1}{n_{M_\nu}}\sum_{\mathcal{P}(J)}^{M!} \sum_{J}^{C^N_M} \prod_{k=1}^{M}\frac{A_{j_k}}{\o_{k,\nu}-A_{j_k}}S_{j_k}^-|0\rangle\label{baes},
\ee
where $A_0=\infty$ by definition. Before writing down the normalization factor $n_{M_\nu}$, let us comment on the last formula. By $|0\rangle$ we mean the fully polarized state $|\Up;\up,\ldots,\up\rangle$, where the symbols $\Up,\Down$ for the central spin and $\up,\down$ for the bath spins are used. In the second step of \refeq{baes}, the product of the previous equation is expanded into a sum over different ordered configurations $J=\{j_1,\ldots,j_M\}$, where $M$ spins are flipped at the  sites $j_{1,\ldots,M}$. The first sum in the final expression of \refeq{baes} comprises the $M!$ permutations of the set $J$. The normalization factor $n_{M_\nu}$ was conjectured by Gaudin \cite{gaubook,gau95} and proved by Sklyanin \cite{skl99} for $h=0$:
\be
n^2_{M_\nu}&=&(-1)^{M} \det \M^{(\nu)}\nn\\
\M_{kk}^{(\nu)}&=&-1-\sum_{j=1}^{N_b}\frac{A_j^2}{(\o_{k,\nu}-A_j)^2}+\sum_{k'\neq k}\frac{2 \,\o_{k',\nu}^2}{(\o_{k,\nu}-\o_{k',\nu})^2}\nn\\
\M_{kk'}^{(\nu)}&=&-\frac{2 \,\o_{k',\nu}^2}{(\o_{k,\nu}-\o_{k',\nu})^2},\; k\neq k'\nn.
\ee
For $M=1$ this is obviously true also for finite $h$. We have checked the validity for $N_b=3,\,M=2$ and finite $h$ as well and thus conjecture that this formula holds for general $N_b,M,h$.

In order to arrive at an expression for $\langle S_0^z\rangle(t)$, we decompose the initial state into Bethe ansatz eigenstates (\ref{baes}). We focus here on the case where the initial state is a  product state. It is denoted by $|L\rangle$, where $L=\l\{\ell_1,\ldots,\ell_{M}\r\}$ is the set of lattice sites with initially flipped spins. 
Let us rewrite Eq.~\refeq{baes} by introducing the $C^N_M\times C^N_M$ matrix $D$ such that
\be
 |M_\nu\rangle =\sum_J^{C_M^N}D_{\nu,J}|J\rangle\,\,,\;\;D_{\nu,J}=\frac{1}{n_{M_\nu}}\sum_{\mathcal{P}(J)}^{M!} \prod_{k=1}^{M}\frac{A_{j_k}}{\o_{k,\nu}-A_{j_k}}\nn,
\ee
where $J$ again denotes the set of locations of flipped spins. From the hermiticity of the Hamiltonian and from the normalization of the eigenstates it follows that $D$ is unitary: $\l[D^{-1}\r]_{\nu,J}=D^*_{J,\nu}$, where $*$ denotes complex conjugation. Thus the time evolution of the initial state reads 
\be
|L\rangle(t)&=&\sum_{\nu,J}D^*_{\nu,L}D_{\nu,J}\,\te^{-\rmi \L_\nu t}|J\rangle\label{lt} .
\ee
We assume that the central spin is {initially} flipped, i.e. $M_b=M-1$. Then $\ell_1\equiv 0$ and the initial configuration is completely determined by the set $L_b=\{\ell_2,\ldots,\ell_M\}$ which contains bath sites only. Ordered sets $J_b$ can be defined analogously. From Eq.~\refeq{lt}, one can infer the reduced density matrix for the central spin
\be
\rho_0(t)&=&\sum_{J_b}^{C_{M_b}^{N_b}}|\al_{J_b}(t)|^2|\!\Down\rangle\langle\Down|\nn\\
& & +\l(1-\sum_{J_b}^{C_{M_b}^{N_b}}|\al_{J_b}(t)|^2\r)|\!\Up\rangle\langle\Up|\label{rho0}
\ee
with
\be
\al_{J_b}(t)&=&\sum_{\nu}D^*_{\nu,0\cup L_b}D_{\nu,0\cup J_b}\te^{\frac{\rmi}{2}t\sum_{k=1}^M\omega_{k,\nu}}\label{aljb}.
\ee
The ordered set $0\cup L_b$ contains the central spin site 0 and $M_b$ bath sites; $0\cup J_b$ is defined similarly. 
We also inserted the eigenvalue \refeq{ev} and dropped an overall phase factor in the last equation. {From Eq.~\refeq{rho0}}, we conclude that 
\be
\langle S_0^z\rangle(t)=\frac12\l(1-2 \sum_{J_b}^{C_{M_b}^{N_b}}|\al_{J_b}(t)|^2\r)\label{sz} .
\ee
Let us pause briefly to comment on the structure of (\ref{sz}) which is the central result of this paper. 
Each $\al_{J_b}(t)$ is a superposition of $C_M^N$ exponentials $\te^{-\rmi t \Lambda_{\nu}}$, with each $\Lambda_{\nu}$ the sum of $M=M_b+1$ Bethe ansatz numbers $\omega_{k,\nu}$, apart from trivial constants and factors, see (\ref{ev}). The expectation value (\ref{sz}) thus contains oscillations with combinations of all these frequencies. 
{It} depends on the initial state through the dependence of $|\al_{J_b}(t)|$ in Eq.~\refeq{aljb} {on $L_b$, the set of initially flipped bath spins}. Let us also comment on the Poincar\'e recurrence time $\tau_P$ after which the system {returns to} the initial state. Generally, BA numbers $\o_{k,\nu}$ are {complex (either real or complex conjugate pairs), such that the eigenvalues $\Lambda_\nu$, Eq.~\refeq{ev}, are irrational numbers. This means that} the system never reaches its initial configuration again, i.e. $\tau_P\to\infty$. In special cases, however, the recurrence time can be finite, for example for homogeneous couplings, $A_j \equiv A$ \cite{bor06}, where an exact solution without the BA is possible. Thus in that particular case, the BA numbers must be rational, which we verified for $M_b=0$, $N$ arbitrary and for $N_b=3$, $M_b=1$.  

%
\section{Special cases}
\label{spec} 
In the homogeneous case\footnote{This case requires a careful treatment of degeneracies that occur in the spectrum and in the BA roots \cite{gau95}.} ($A_j\equiv A\, \forall\, j$) the Hamiltonian can be diagonalized directly, i.e. without employing the Bethe ansatz \cite{bor06}. We have checked that 
for $N_b$ arbitrary, $M_b=0$ and for $N_b=3$, $M_b=1$, Eq.~\refeq{sz} yields the same results as those obtained in \cite{bor06}. 
\subsection{Fully polarized bath: Analytical results}
Let us now concentrate on the inhomogeneous case with arbitrary $N_b$ and a fully polarized bath ($M_b=0$). The initial state {then} is $|\!\Down;\up\ldots\up\rangle$. The sum in Eq.~\refeq{aljb} {simplifies to}
\be
\al(t)&=&\sum_{\nu=1}^N\l[1+\sum_{j=1}^{N_b}\frac{A_j^2}{(A_j-\omega_\nu)^2}\r]^{-1}\te^{\frac{\rmi}{2}\omega_\nu t}\label{al1}, 
\ee
where the frequencies $\omega_\nu$ are solutions of the single BA equation \footnote{It is worthwhile noting that these equations can also be obtained by the ansatz made in \cite{kha02,kha03}. In that approach, the BA numbers are the poles of the Laplace transform of $\al(t)$. In contrast to \cite{kha02,kha03}, where the {many}-particle limit is taken in the Laplace domain, here this limit is taken in the time-domain.}
\be
\sum_{j=1}^{N_b}\frac{A_j}{A_j-\omega_\nu}=-1-\frac{2h}{\omega_\nu}\label{ba} ,
\ee
and $\langle S_0^z\rangle(t)$ is given by
\be
\langle S_0^z\rangle(t)=\frac12\l(1-2 |\al(t)|^2\r)\label{sfromal}\,.
\ee
Before solving these equations numerically, we first extract the qualitative behaviour of $\langle S_0^z\rangle(t)$ in the limit of large $N_b$. Therefore, the $N$ roots $\omega_\nu$ of (\ref{ba}) have to be found. {In all that follows, {we restrict ourselves to the physically relevant case $A_j>0 \,\forall j$} and assume the couplings to be distributed such that for large particle numbers, $N_b \bar A \sim N_b $ and $\sum A_j^2 \sim N_b$.} 

\subsubsection{Weak field}

Let us first consider a weak magnetic field, {$|h|/N\ll \mbox{inf}_j A_j $}. Then 
\be
\o_1=-2h/N+\Or(h^2/N^2)\label{om1}
\ee
is one solution, tending to $0$ as $h\to 0$. Another solution is found at 
\be
\o_2=N_b \bar A-2 h+\Or(1)\label{om2}.
\ee
By {sketching a graphical solution} of Eq.~\refeq{ba} one sees that 
each of the remaining $N-2$ solutions is located between two consecutive $A_j$.

For the sake of simplicity, we write down $\al(t)$ first for $h=0$,
and calculate the amplitudes in (\ref{al1}) only to order $\Or(1/N_b)$. 
\be
\lefteqn{\al(t)=\l(1{-}\frac{\sum_{j=1}^{N_b} A_j^2}{(N_b \bar A)^2}\r) \te^{\frac{\rmi}{2} N_b \bar A t}+\frac{1}{N_b}}\nn\\
& &+ \sum_{\nu=3}^N\l[1+\sum_{j=1}^{N_b}\frac{A_j^2}{(A_j-\omega_\nu)^2}\r]^{-1}\te^{\frac{\rmi}{2}\omega_\nu t}\label{alt2}\\
& &+{\Or\l(1/N_b^{2}\r)}\,.\nn
\ee
To obtain a crude approximation of the last sum, we recall that each $\omega_\nu$ is located between two consecutive $A_j$. Let $0< A_j\leq \mathcal A$, where $\mathcal A$ is independent of $N$, and let the $A_j$ be distributed smoothly between $0$ and $\mathcal A$. Then we estimate the difference between two consecutive $A_j$ as $\sim \mathcal A/N_b$, such that those $A_j$ which are closest to $\omega_\nu$ in the last bracket in \refeq{alt2} dominate and {yield a contribution} $\sim \mathcal A^2/N_b^2$ to each term of the sum over the $\omega_\nu$. We assume here that the number of these terms does not scale with $N_b$ such that the whole sum behaves as
\be
\sum_{j=1}^{N_b}\frac{A_j^2}{(A_j-\omega_\nu)^2}\sim N_b^2\label{sum}\, .
\ee  
Obviously, we cannot perform the thermodynamic limit here directly since then the sum (\ref{sum}) would diverge. Consequently the third 
{term} in \refeq{alt2} would vanish, as would the second term, resulting in trivial dynamics. (Note that $|\alpha(t)|^2$ would equal unity in that case.)
Thus $\alpha(t)$ above has to be evaluated for large, but still finite $N_b$, avoiding singularities that occur for $N_b\to \infty$. 

Therefore the number of particles is kept finite, and, following Eq.~\refeq{sum}, we replace $\l[1+\sum_{j=1}^{N_b}\frac{A_j^2}{(A_j-\omega_\nu)^2}\r]^{-1} = \gamma_\nu/N_b^2$, with certain constants $\gamma_\nu$. Then
\be
\lefteqn{\sum_{\nu=3}^N\!\l[1+\sum_{j=1}^{N_b}\frac{A_j^2}{(A_j-\omega_\nu)^2}\r]^{-1}\!\te^{\frac{\rmi}{2}\omega_\nu t}}\nn\\
& &= \frac{1}{N_b^2}\sum_{\nu=3}^{N} \gamma_\nu\te^{\frac{\rmi}{2}\o_\nu t}\label{app2}.
\ee
{We will now show that this sum describes an amplitude-modulated} oscillation where the envelope has its first zero around 
\be
t_a:=4\pi/\mathcal A\label{ta}\; .
\ee
For a large number of particles, {we} approximate the last sum in Eq.~\refeq{app2} {by an integral. To leading order this yields}
\be
\sum_{\nu=3}^{N} \gamma_\nu\te^{\frac{\rmi}{2}\o_\nu t}\approx N_b \,\int_{0}^{\mathcal A} f(\omega) \te^{\frac{\rmi}{2} \omega t} \rho(\omega) \d \omega+\Or(1), \nn
\ee
where $f(\omega)$  is the continuum limit of the $\gamma_\nu$ in Eq.~\refeq{app2} and $\rho(\omega)$ is the root density. Proceeding further, we can write
\be
\lefteqn{\frac{1}{N_b}\sum_{\nu=3}^{N} \gamma_\nu\te^{\frac{\rmi}{2}\o_\nu t}}\nn\\
&\approx& \te^{\rmi b(\mathcal{A}) t} \,\int_{-\mathcal{A}/2}^{\mathcal{A}/2} f(\omega+\mathcal{A}/2)\te^{\frac{\rmi}{2}\omega t}\rho(\omega+\mathcal{A}/2)\,\d \omega\nn\\
&=:&\te^{\rmi b(\mathcal{A}) t} F_{\mathcal A} (t) \label{slot},
\ee
where the constant $b(\mathcal A)$ in the phase factor is chosen such that $F_{\mathcal A}$ is real (if $f(\omega)$ and $\rho(\omega)$ are constants, then $b(\mathcal A)=\mathcal A/4$). The integral is the Fourier transform of a {band-limited function of bandwidth $\mathcal A$. The frequency cutoff at $\pm \mathcal A/2$ leads} to a decaying oscillation with a period {$8 \pi/\mathcal A$}, which reaches its first zero around $t_a$. By inserting the result \refeq{slot} into Eq.~\refeq{alt2}, we evaluate $|\al(t)|^2$ including orders $1,1/N_b$:
\be
\lefteqn{|\al(t)|^2\approx1-\frac{2}{(N_b \bar A)^2}\sum_{j=1}^{N_b}A_j^2}\label{altf}\\
&+&\frac{2}{N_b}\l\{ \cos\l[\frac{N_b \bar A}{2}t\r]+F_{\mathcal A}(t)\cos\l[\l(\frac{N_b \bar A}{2}-b(\mathcal A)\r)t\r] \r\}\nn.
\ee
Thus the qualitative behaviour of $|\al(t)|^2$ is as follows. At $t=0$, the initial condition is $|\al(0)|^2=1$. After that, $|\alpha(t)|^2$ displays fast oscillations  around the mean value 
\be
\overline{|\al|^2}=1-\frac{2}{(N_b \bar A)^2}\sum_{j=1}^{N_b}A_j^2\label{mval}
\ee
with the leading frequency 
\be
\omega_l:=\frac{N_b \bar A}{2}.\label{oml}
\ee
These oscillations are given by the two cosine terms in Eq.~\refeq{altf}. Both terms have amplitudes of order $\Or(1/N_b)$. But whereas the amplitude of the first $\cos$-term is constant, the amplitude of the second one depends on time. Especially, if $\mathcal  F_{\mathcal A}(0)>1$, the latter term will dominate initially, decaying $\sim t^2$ for short times $t\ll t_a$. As stated above, $\mathcal F_{\mathcal A}(t)$ decays and it is certainly dominated by the other cosine term at times $t\approx t_a$. As the exact numerical evaluation below (see Fig. \ref{fig1}) shows for both a uniform and a nonuniform distribution of the couplings, the constant amplitude term also dominates for all times $t\gtrsim t_a$. 

{In any case, $|\al(t)|^2$ does not show any long-term decay, and consequently neither does $\langle S_0^z\rangle (t)$. We rather expect 
\be
\lefteqn{|\al(t)|^2}\nn\\
&&\!\!\!\approx  1-\frac{2}{(N_b \bar A)^2}\sum_{j=1}^{N_b}A_j^2 +\frac{c_1}{N_b} \cos\l[\l(\frac{N_b \bar A}{2}-h\r)t\r]\label{altf2} 
\ee  
for long times $t\gg t_a$, with an undetermined amplitude $c_1$ of order $\Or(1)$. Although the {calculations leading to the results (\ref{altf},\ref{altf2}) certainly lack mathematical rigor}}{, they are confirmed in the next section by a numerical evaluation of the exact Eqs.~(\ref{al1}, \ref{ba}).} 

{Note that the last sum in Eq.~\refeq{alt2} is completely due to the inhomogeneities of the couplings: If all couplings are equal, the Bethe Ansatz equation \refeq{ba} is a quadratic equation in $\omega$ with the two solutions $\omega_{1,2}$ (the exact solutions in that case are given below). Summarizing, the non-uniformity of the couplings causes an initial decay of the amplitude until a time $\sim t_a$. For longer times, the amplitude does not change any more and $\langle S_0^z \rangle(t)$ keeps oscillating with an amplitude of order $\Or(1/N_b)$ about a mean value $-\frac12 + \Or(1/N_b)$. 
{This does not agree with \cite{kha02,kha03}, where the amplitude of oscillations is predicted to vanish in the long-time limit.} The leading frequency of the oscillation is $\omega_l$ (\ref{oml}), in agreement with \cite{kha02,kha03}. A small positive magnetic field causes a slow-down of the oscillation, as can be seen from Eq.~(\ref{om2}).}

\subsubsection{Strong field}
     
We now discuss the case of a strong field $|h|\gg N_b$. 
Here the limiting values of the two solutions $\omega_{1,2}$ considered above in the weak field case, Eqs.~(\ref{om1},\ref{om2}), are given by 
\be
\omega_1=-2 h+N_b \bar A+\Or(N_b/h)\label{om1h} \\
\omega_2= A_{1}(1+A_{1}/(2h))+\Or(1/h^2)\label{om2h}.
\ee
{The remaining BA} roots are found at 
\be
\omega_{\nu}=A_{\nu-1}(1+A_{\nu-1}/(2h))+\Or(1/h^2),\; \nu=3,\ldots,N.\label{allomh}
\ee
{Thus the magnetic field shifts the roots along the real axis. (See Fig.~\ref{figba} for an example.)}

{Eqs.~(\ref{om1h},\ref{om2h},\ref{allomh}) are inserted into Eq.~\refeq{al1}, keeping only the leading non-trivial contribution in the amplitudes:} 
\be
\lefteqn{\al(t)\approx  \l(1-\frac{1}{4 h^2}\sum_{j=1}^{N_b} A_j^2\r) \te^{-\rmi \l(h- N_b \bar A/2\r)t}}\nn\\
&&+\sum_{j=1}^{N_b}\frac{A_j^2}{4 h^2}\,\te^{\frac{\rmi}{2}(A_j+{A_j^2/(2 h)}) t}\nn.
\ee
Arguing  as in the zero-field case after Eq.~\refeq{alt2}, we find
\be
\lefteqn{|\al(t)|^2\approx 1-\frac{1}{2 h^2}\sum_{j=1}^{N_b} A_j^2}\nn\\
& &+\frac{2N_b}{h^2}\l[F_{\mathcal A}^{(h)}(t)\r] \cos\l[\l(h-N_b \bar A/2+b_h(\mathcal{A})\r)t\r]\label{lh}, \ee
\tbl {} 
with 
\be 
{\te^{\rmi t b_h(\mathcal{A})}} F_{\mathcal A}^{(h)}(t)=\int_{{0}}^{{\mathcal A}} A^2 \te^{\frac{\rmi}{2} (A+{A^2/(2h)}) t} \tilde \rho(A) \d A \nn.
\ee
Here $\tilde \rho(A)$ is the density of the coupling constants $A_k$, and $b_h(\mathcal{A})$ in Eq.~\refeq{lh} is chosen such that $F_{\mathcal A}^{(h)}(t)$ is real. Note that in this case, the result depends only on the (known) couplings $A_k$. It is expected to be valid for times as long as the $\omega_\nu$ can be approximated by Eqs. (\ref{om1h},\ref{om2h},\ref{allomh}), i.e., $t\ll h/N_b$. 
The enveloping function $F_{\mathcal A}^{(h)}(t)$ again describes a decreasing oscillation. For times $t\gg h{/N_b}$, the replacement {of the} $\omega_\nu$ by the $A_{\nu-1}$ is no longer valid, and $|\al(t)|^2$ oscillates around a mean value $1-\frac{1}{2 h^2}\sum_{j=1}^{N_b} A_j^2$ with an amplitude $\sim N_b/h^2$ and frequency $\sim |h-N_b \bar A/2|$. Note that increasing a strong  field suppresses the amplitude further and raises
 the frequency. Combining this with the findings for the small field case, we conclude that for $h>0$, there should be a resonance region around a field $h_r$ where the amplitude is maximal and the frequency minimal.

\subsubsection{Resonance field}
To study that region, it is instructive to consider homogeneous couplings first, $A_j=A \forall j$. {In that case the BA equation \refeq{ba} becomes a quadratic equation with the two solutions $\omega_{1,2}$ given below. The remaining solutions $\omega_{\nu=3,\ldots,N}$ all tend to  $\omega_{\nu}=A$ when $A_j \to A \forall j$, as can be shown using  techniques from \cite{gau95}. The contributions to $\alpha(t)$, {Eq.}~(\ref{al1}), corresponding to those roots all vanish and we end up with}
\begin{widetext}
\be
|\al(t)|^2&=&\l[\frac{A}{\omega_1-\omega_2}\r]^2\l\{\l(\frac{\omega_1}{A}-1\r)^2+\l(\frac{\omega_2}{A}-1\r)^2-2\l(\frac{\omega_1}{A}-1\r)\l(\frac{\omega_2}{A}-1\r)\cos\l[\l(\omega_1-\omega_2\r)t/2\r]\r\}\nn\\
\omega_{1,2}&=&\frac12\l(AN-2 h \mp \sqrt{8Ah+(A N-2h)^2}\r)\nn\\
\langle S_0^z\rangle(t)&=&-\frac{1}{2\l((AN-2h)^2+8 A h\r)}\l(\l(AN-2(A+h)\r)^2+4A^2(N-1) \cos\l[\frac12\sqrt{(AN-2h)^2+8 A h}t\r]\r)\nn\,.
\ee
\end{widetext}
For $h=A(N-2)/2=:h_r$, this results in $S_0^z(t)=-\frac12 \cos\l[A \sqrt{N_b}\, t\r]$ and 
\be
\omega_{1,2}=A\l(-1\mp \sqrt{N_b}\r)\label{om12spec}.
\ee
Thus if $h$ equals the $\Or(N_b)$-magnetization of the bath, the central spin resonates with maximal amplitude and the two roots {$\omega_{1,2}$} have equal absolute values in the limit of a large particle number. The width of the resonance is estimated by considering the time averaged value $\overline{\langle S_0^z\rangle}$ as a function of $h$. From this one deduces a width $\sim \sqrt{N_b}$. This behaviour was also found in \cite{kha02,kha03} for the inhomogeneous case, which we turn to now. 

{By analogy to the homogeneous case we expect resonance to occur at 
\be
h_r=N_b\bar A/2\label{resfield}.
\ee
>From the discussion of the large- and small-$h$ limits above we expect that at least one $|\omega_{\nu}|$ will be much larger than all $A_j$. Expanding the BA equation (\ref{ba}) to second order in $\omega_{\nu}^{-1}$ we obtain a quadratic equation for the extremal $\omega_{\nu}$:
\be
 \frac{\sum_{j=1}^{N_b} A_j}{\omega_\nu} +  \frac{\sum_{j=1}^{N_b} A_j^2}{\omega_\nu^2} =1+\frac{2 h}{\omega_\nu}.
\ee
For the expected resonance field value (\ref{resfield}) this leads to
\be
\omega_{1,2}^2= \sum_{j=1}^{N_b} A_j^2+\Or(1)\label{om12}.
\ee 
Hence the two extremal BA roots $\omega_{1,2}$, for which limiting values for small and large fields have been given in Eqs.~(\ref{om1},\ref{om2}) and (\ref{om1h},\ref{om2h}), respectively, have equal absolute values, and the resonant behaviour is determined (to leading order) by}
\be
|\al(t)|^2=\frac{1}{2}\l[1+\cos\l(\omega_{1,2}\,t\r)\r]\label{omapp}.
\ee
 The width of the resonance is still $\sim \sqrt{N_b}$, in agreement with \cite{kha02,kha03}. 

\begin{figure}
\begin{center}
\includegraphics*[scale=0.3]{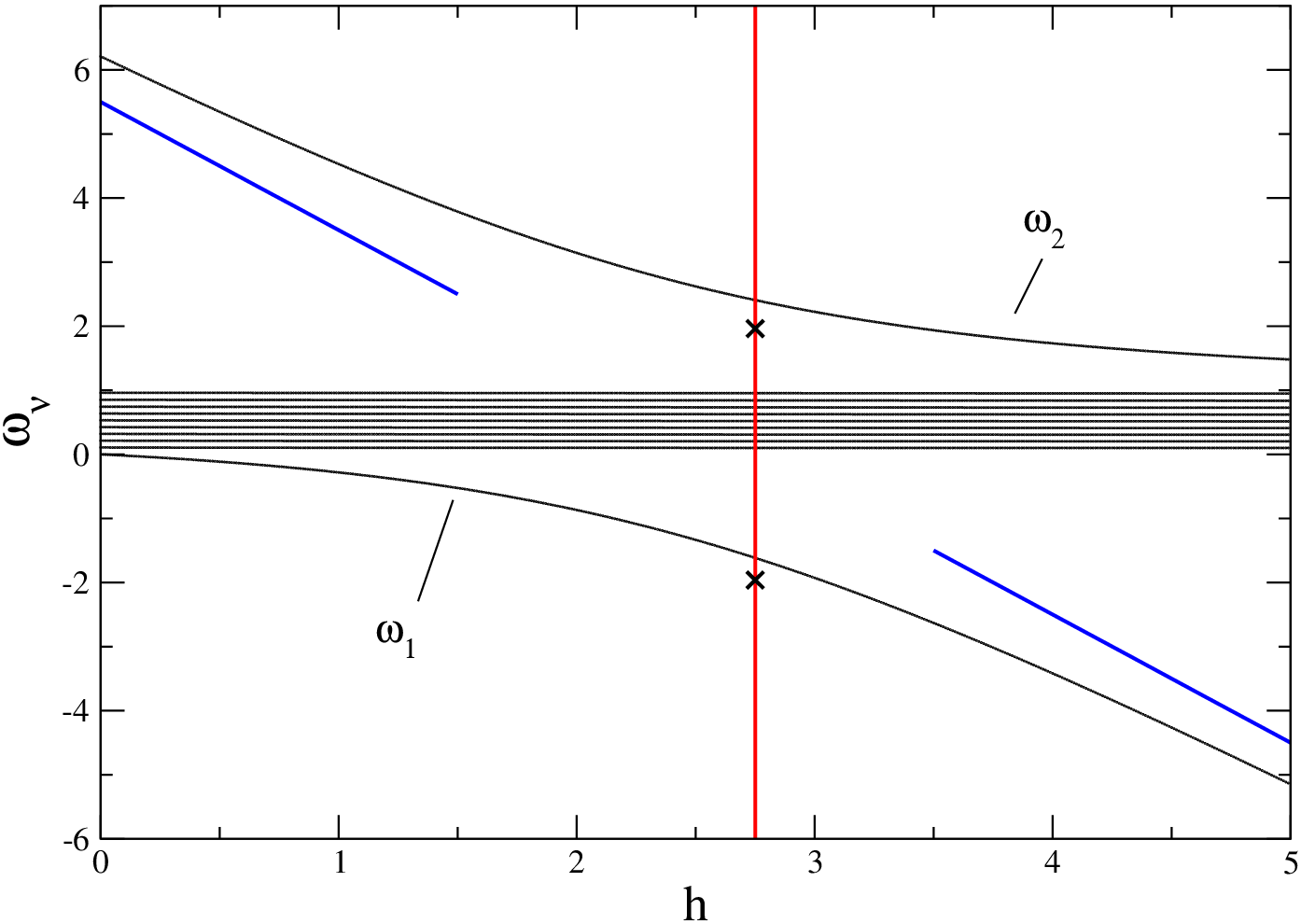}
\caption{The locations of the $N$ {Bethe ansatz numbers} $\omega_\nu$ for
  $N=11$ and a uniform distribution of the couplings between $0$ and $1$. The
  (blue) straight diagonal lines to the left and right are the small- and
  large-field approximations Eqs.~(\ref{om2},\ref{om1h}). The (red) vertical line denotes the resonance field $h_r$, Eq.~\refeq{resfield}, and the two crosses are the approximate values for $\omega_{1,2}$ from Eq.~\refeq{om12}.} 
\label{figba}
\end{center}
\end{figure}

{We illustrate the evolution of the BA roots with $h$ by solving Eq.~\refeq{ba} numerically for $N_b=10$ and $A_j=(11-j)/10$, $j=1,\ldots,10$ and  $h$ between 0 and 5. The results are depicted in Fig.~\ref{figba}. As expected, the ``inner roots'' $\omega_{3,\ldots,N}$ are always located between two successive $A_j$, whereas the location of the ``outer roots'' $\omega_{1,2}$ changes significantly with $h$.}
\subsection{Fully polarized bath: Numerical results}
In the following, we show exact results for $\langle S_0^z\rangle(t)$ with $N_b=30$ bath spins, obtained by numerically solving the BA equation \refeq{ba}, and compare those {results with the analytic approximations just derived}. Note that the features {of those approximations} are universal in the sense that they depend on the distribution of the $A_j$ only through the mean value $\bar A$. 
 We consider two different choices of distributing the $A_j$ {between zero and $\mathcal{A}=1$; a nonuniform distribution and a uniform one:}
\be
A_j&=&\exp\l[-(j/a)^2\r]\;\;\mbox{with  } a=N_b/2\label{gaudis}\\
A_j&=&(N_b+1-j)/N_b\label{lindis}\;
\ee
for $j=1,\ldots,N_b$.

\begin{figure}
\begin{center}
\includegraphics*[scale=0.3]{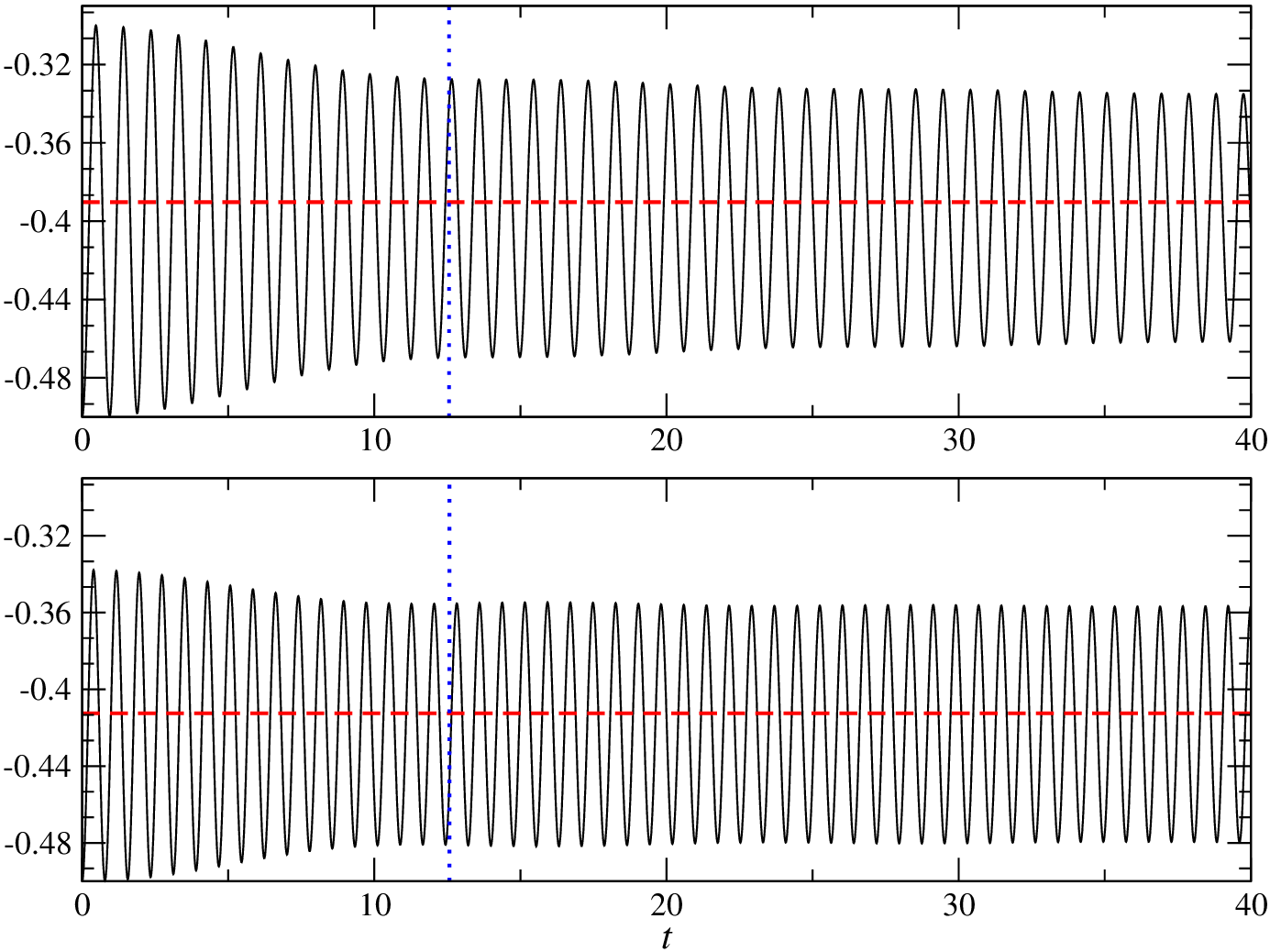}
\caption{$\langle S_0^z\rangle(t)$ for $h=0${, $N_b=30$ bath spins} and a {non-uniform (upper panel) and uniform} distribution (lower panel) according to Eqs.~\refeq{gaudis},\refeq{lindis} respectively. Red dashed lines denote the mean value \refeq{mv}, and blue dotted lines the time scale $t_a$, defined in Eq.~\refeq{ta}. {The initial decay of the envelope} is $\sim t^2$, as expected from Eq.~\refeq{altf}.} 
\label{fig1}
\end{center}
\end{figure}

Figure \ref{fig1} shows the time evolution of $\langle S_0^z\rangle$ for {zero field}, $h=0$. The initial decay $\sim t^2$ {of the envelope}, the time scale $t_a$ where {that} decay stops, the frequency and the mean value around which {$\langle S_0^z \rangle (t)$ oscillates} agree well with the exact numerics. We recall that from combining Eqs.~(\ref{sfromal},\ref{mval}), the mean value follows
\be
\overline{\langle S_0^z\rangle} &=& -\frac{1}{2}\l(1-\frac{4}{(N_b \bar A)^2}\sum_{j=1}^{N_b} A_j^2\r)\label{mv}\,,
\ee
whereas the leading (i.e. the fast) oscillation is given by $\omega_l$, Eq.~\refeq{oml}. 
{In order to compare that (approximate) value with the {exact} numerics we counted the oscillations within the time interval shown. The numerical values thus determined are
 $\omega_{l,n}=6.7$, ($\omega_{l,n}=8.1$) for the nonuniform (uniform) distributions, in good agreement with the analytical predictions $\omega_l=\frac{N_b \bar A}{2}=6.4$ (7.8).}

\begin{figure}
\begin{center}
\includegraphics*[scale=0.3]{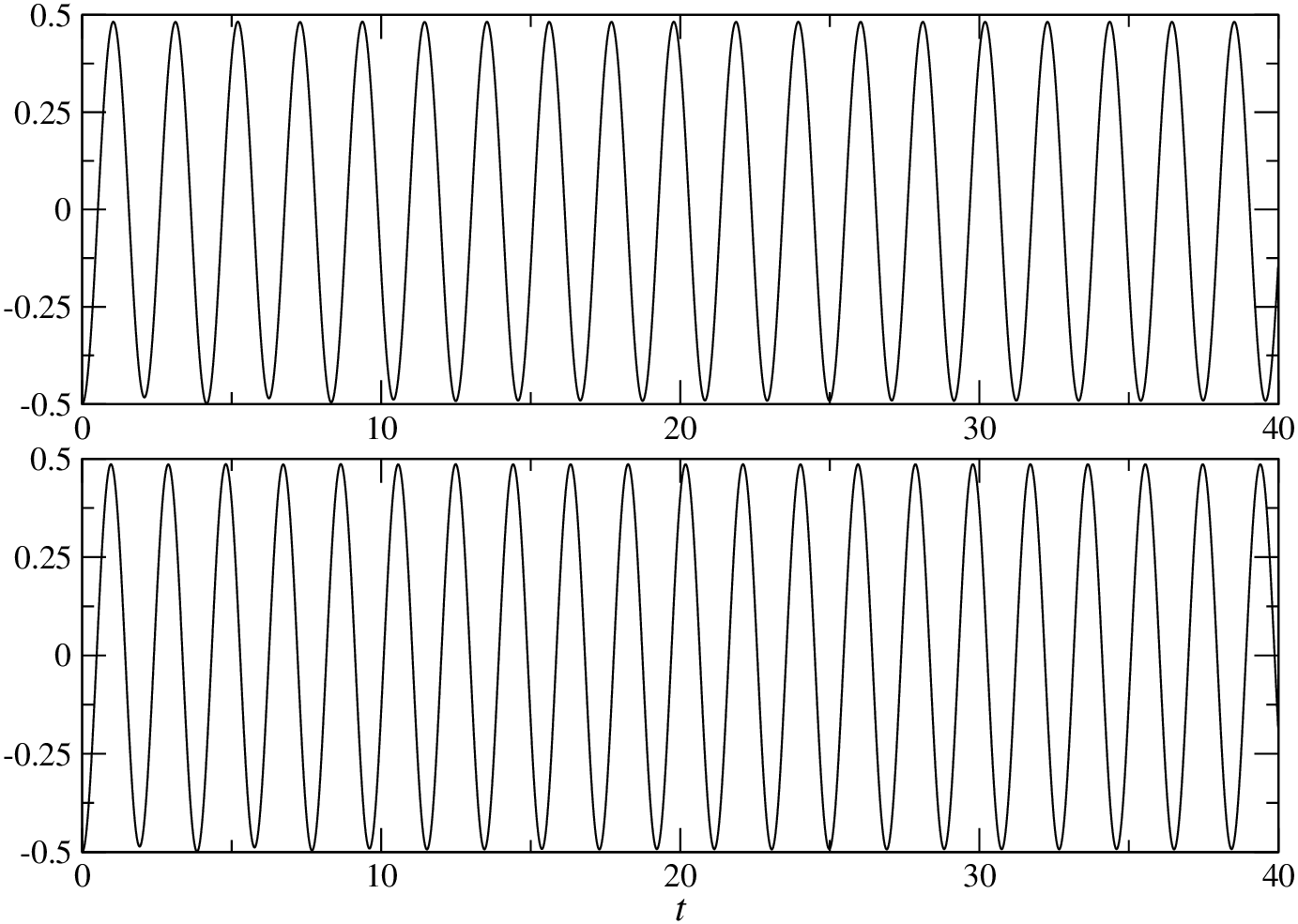}
\caption{$\langle S_0^z\rangle(t)$ for $h=h_r${, $N_b=30$ bath spins} (cf. Eq.~\refeq{resfield}) and a nonuniform (upper panel) and uniform distribution (lower panel) according to Eqs.~\refeq{gaudis},\refeq{lindis} respectively. In this case, $h_r=6.37$ for the nonuniform and $h_r=7.75$ for the uniform distribution.} 
\label{fig2}
\end{center}
\end{figure}

The resonance case $h=h_r$ is shown in Fig.~\ref{fig2}. The oscillation is between $\pm 1/2$, as expected {from Eqs. (\ref{sfromal},\ref{omapp})}. The numerical values for the frequencies
$\omega_{l,n}=3.0$ ($\omega_{l,n}=3.3$) for the nonuniform (uniform) distribution compare well with the approximate values derived from Eqs.~(\ref{om12},\ref{omapp}), $\sqrt{\sum_{j=1}^{N_b} A_j^2}=3.0$ (3.2). 

For the case of a large field, we checked that the initial decay is $\sim t^2$, in agreement with Eq.~\refeq{lh}. Furthermore, the approximate mean value $\overline{\langle S_0^z \rangle}=-1/2(1-\frac{1}{h^2}\sum_{j=1}^{N_b} A_j^2)$ coincides with numerical results as well. 
\section{Conclusion and outlook}
The approximate analytical and {exact} numerical evaluations of the {general} formula for the time evolution of the central spin polarization $\langle S_0^z\rangle$ for an initially fully polarized spin bath show that {in this case}, an inhomogeneous broadening of the Heisenberg couplings leads to decoherence only initially for short times. {As exemplified by Fig. \ref{fig1},}
this decoherence process is far from complete and does not suppress the oscillations of the central spin completely at long times. 

{Our previous work \cite{bor06} on the central spin model with homogeneous couplings shows  that an initially unentangled spin  bath supports complete decoherence, if the magnetization of the bath is zero or small. In that case  $\langle S_0^z\rangle(t)$ decays to zero within a decoherence time $\tau_d\sim 1/\sqrt{N_b}$. However, at a later time $\tau_P=\Or(1)$ (the Poincar\'{e} recurrence time)  $\langle S_0^z\rangle(t)$ shows a complete revival due to the commensurate energy spectrum of the homogeneous model.  In the inhomogeneous model we expect a divergent Poincar{\'e} recurrence time, $\tau_P\to \infty$.}

In contrast {to that} an entangled bath state with zero or small magnetization may lead to persistent oscillations of $\langle S_0^z\rangle(t)$ at maximum amplitude in the homogeneous model. We thus see that in the homogeneous model initial bath states with zero magnetization but different degrees of entanglement may lead to very different long-time behavior{s}.
This observation is
consistent with \cite{daw05,ros06}, where it was argued that entanglement in the bath protects the central spin from decohering. If this scenario is generally valid remains to be studied further.

{Another} very interesting question is {to what extent} non-uniformity of the couplings  affects the decoherence of the central spin for an unmagnetized bath{, as opposed to the completely magnetized bath studied in the present paper}. Although this question has been addressed with perturbative methods before \cite{kha02,kha03}, the analysis of the {general solution derived in the present paper} is expected to yield {deeper and more quantitative}  insights.  
\section*{Acknowledgments}
We thank M.T. Batchelor, W.A. Coish, X.-W. Guan and D. Loss for helpful discussions. This work has been supported by the German Research Council (DFG) under grant number BO2538/1-1. 
\bibliography{gaudin_inh} 

\begin{thebibliography}{20}
\expandafter\ifx\csname natexlab\endcsname\relax\def\natexlab#1{#1}\fi
\expandafter\ifx\csname bibnamefont\endcsname\relax
  \def\bibnamefont#1{#1}\fi
\expandafter\ifx\csname bibfnamefont\endcsname\relax
  \def\bibfnamefont#1{#1}\fi
\expandafter\ifx\csname citenamefont\endcsname\relax
  \def\citenamefont#1{#1}\fi
\expandafter\ifx\csname url\endcsname\relax
  \def\url#1{\texttt{#1}}\fi
\expandafter\ifx\csname urlprefix\endcsname\relax\def\urlprefix{URL }\fi
\providecommand{\bibinfo}[2]{#2}
\providecommand{\eprint}[2][]{\url{#2}}

\bibitem[{\citenamefont{Gaudin}(1976)}]{gau76}
\bibinfo{author}{\bibfnamefont{M.}~\bibnamefont{Gaudin}}, \bibinfo{journal}{J.
  Physique} \textbf{\bibinfo{volume}{37}}, \bibinfo{pages}{1087}
  (\bibinfo{year}{1976}).

\bibitem[{\citenamefont{Gaudin}(1983)}]{gaubook}
\bibinfo{author}{\bibfnamefont{M.}~\bibnamefont{Gaudin}},
  \emph{\bibinfo{title}{La fonction d'onde de Bethe}}
  (\bibinfo{publisher}{Masson}, \bibinfo{year}{1983}).

\bibitem[{\citenamefont{Schliemann et~al.}(2002)\citenamefont{Schliemann,
  Khaetskii, and Loss}}]{sch02}
\bibinfo{author}{\bibfnamefont{J.}~\bibnamefont{Schliemann}},
  \bibinfo{author}{\bibfnamefont{A.}~\bibnamefont{Khaetskii}},
  \bibnamefont{and} \bibinfo{author}{\bibfnamefont{D.}~\bibnamefont{Loss}},
  \bibinfo{journal}{Phys. Rev. B} \textbf{\bibinfo{volume}{66}},
  \bibinfo{pages}{245303} (\bibinfo{year}{2002}).

\bibitem[{\citenamefont{Schliemann et~al.}(2003)\citenamefont{Schliemann,
  Khaetskii, and Loss}}]{sch03}
\bibinfo{author}{\bibfnamefont{J.}~\bibnamefont{Schliemann}},
  \bibinfo{author}{\bibfnamefont{A.}~\bibnamefont{Khaetskii}},
  \bibnamefont{and} \bibinfo{author}{\bibfnamefont{D.}~\bibnamefont{Loss}},
  \bibinfo{journal}{J. Phys.: Cond. Mat.} \textbf{\bibinfo{volume}{15}},
  \bibinfo{pages}{R1809} (\bibinfo{year}{2003}).

\bibitem[{\citenamefont{Dobrovitski and De~Raedt}(2003)}]{dob03}
\bibinfo{author}{\bibfnamefont{V.}~\bibnamefont{Dobrovitski}} \bibnamefont{and}
  \bibinfo{author}{\bibfnamefont{H.}~\bibnamefont{De~Raedt}},
  \bibinfo{journal}{Phys. Rev. E} \textbf{\bibinfo{volume}{67}},
  \bibinfo{pages}{056702} (\bibinfo{year}{2003}).

\bibitem[{\citenamefont{Semenov and Kim}(2003)}]{sem03}
\bibinfo{author}{\bibfnamefont{Y.}~\bibnamefont{Semenov}} \bibnamefont{and}
  \bibinfo{author}{\bibfnamefont{K.}~\bibnamefont{Kim}},
  \bibinfo{journal}{Phys. Rev. B} \textbf{\bibinfo{volume}{67}},
  \bibinfo{pages}{073301} (\bibinfo{year}{2003}).

\bibitem[{\citenamefont{Khaetskii et~al.}(2002)\citenamefont{Khaetskii, Loss,
  and Glazman}}]{kha02}
\bibinfo{author}{\bibfnamefont{A.}~\bibnamefont{Khaetskii}},
  \bibinfo{author}{\bibfnamefont{D.}~\bibnamefont{Loss}}, \bibnamefont{and}
  \bibinfo{author}{\bibfnamefont{L.}~\bibnamefont{Glazman}},
  \bibinfo{journal}{Phys. Rev. Lett.} \textbf{\bibinfo{volume}{88}},
  \bibinfo{pages}{186802} (\bibinfo{year}{2002}).

\bibitem[{\citenamefont{Khaetskii et~al.}(2003)\citenamefont{Khaetskii, Loss,
  and Glazman}}]{kha03}
\bibinfo{author}{\bibfnamefont{A.}~\bibnamefont{Khaetskii}},
  \bibinfo{author}{\bibfnamefont{D.}~\bibnamefont{Loss}}, \bibnamefont{and}
  \bibinfo{author}{\bibfnamefont{L.}~\bibnamefont{Glazman}},
  \bibinfo{journal}{Phys. Rev. B} \textbf{\bibinfo{volume}{67}},
  \bibinfo{pages}{195329} (\bibinfo{year}{2003}).

\bibitem[{\citenamefont{Deng and Xuedong}(2006)}]{den06}
\bibinfo{author}{\bibfnamefont{C.}~\bibnamefont{Deng}} \bibnamefont{and}
  \bibinfo{author}{\bibfnamefont{H.}~\bibnamefont{Xuedong}},
  \bibinfo{journal}{Phys. Rev. B} \textbf{\bibinfo{volume}{73}},
  \bibinfo{pages}{241303(R)} (\bibinfo{year}{2006}).

\bibitem[{\citenamefont{Al-Hassanieh et~al.}(2006)\citenamefont{Al-Hassanieh,
  Dobrovitski, Dagotto, and Harmon}}]{has06}
\bibinfo{author}{\bibfnamefont{K.}~\bibnamefont{Al-Hassanieh}},
  \bibinfo{author}{\bibfnamefont{V.}~\bibnamefont{Dobrovitski}},
  \bibinfo{author}{\bibfnamefont{E.}~\bibnamefont{Dagotto}}, \bibnamefont{and}
  \bibinfo{author}{\bibfnamefont{B.}~\bibnamefont{Harmon}},
  \bibinfo{journal}{Phys. Rev. Lett.} \textbf{\bibinfo{volume}{97}},
  \bibinfo{pages}{037204} (\bibinfo{year}{2006}).

\bibitem[{\citenamefont{Coish et~al.}(2007)\citenamefont{Coish, Yuzbashyan,
  Altshuler, and Loss}}]{coi06}
\bibinfo{author}{\bibfnamefont{W.}~\bibnamefont{Coish}},
  \bibinfo{author}{\bibfnamefont{E.}~\bibnamefont{Yuzbashyan}},
  \bibinfo{author}{\bibfnamefont{B.}~\bibnamefont{Altshuler}},
  \bibnamefont{and} \bibinfo{author}{\bibfnamefont{D.}~\bibnamefont{Loss}},
  \bibinfo{journal}{J. Appl. Phys.} \textbf{\bibinfo{volume}{101}},
  \bibinfo{pages}{081715} (\bibinfo{year}{2007}).

\bibitem[{\citenamefont{Loss and DiVincenzo}(1998)}]{los98}
\bibinfo{author}{\bibfnamefont{D.}~\bibnamefont{Loss}} \bibnamefont{and}
  \bibinfo{author}{\bibfnamefont{D.}~\bibnamefont{DiVincenzo}},
  \bibinfo{journal}{Phys. Rev. A} \textbf{\bibinfo{volume}{57}},
  \bibinfo{pages}{120} (\bibinfo{year}{1998}).

\bibitem[{\citenamefont{Coish and Loss}(2004)}]{coi04}
\bibinfo{author}{\bibfnamefont{W.}~\bibnamefont{Coish}} \bibnamefont{and}
  \bibinfo{author}{\bibfnamefont{D.}~\bibnamefont{Loss}},
  \bibinfo{journal}{Phys. Rev. B} \textbf{\bibinfo{volume}{70}},
  \bibinfo{pages}{195340} (\bibinfo{year}{2004}).

\bibitem[{\citenamefont{Sklyanin}(1989)}]{skl89}
\bibinfo{author}{\bibfnamefont{E.}~\bibnamefont{Sklyanin}},
  \bibinfo{journal}{J. Sov. Math.} \textbf{\bibinfo{volume}{47}},
  \bibinfo{pages}{2473} (\bibinfo{year}{1989}).

\bibitem[{\citenamefont{Gaudin}(1995)}]{gau95}
\bibinfo{author}{\bibfnamefont{M.}~\bibnamefont{Gaudin}}, in
  \emph{\bibinfo{booktitle}{Travaux de Michel Gaudin, {M}od\`eles exactement
  r\'esolus}} (\bibinfo{publisher}{Les Editions de Physique},
  \bibinfo{year}{1995}), p. \bibinfo{pages}{247}.

\bibitem[{\citenamefont{Yuzbashyan et~al.}(2005)\citenamefont{Yuzbashyan,
  Altshuler, Kuznetsov, and Enolskii}}]{yuz05b}
\bibinfo{author}{\bibfnamefont{E.}~\bibnamefont{Yuzbashyan}},
  \bibinfo{author}{\bibfnamefont{B.}~\bibnamefont{Altshuler}},
  \bibinfo{author}{\bibfnamefont{V.}~\bibnamefont{Kuznetsov}},
  \bibnamefont{and} \bibinfo{author}{\bibfnamefont{V.}~\bibnamefont{Enolskii}},
  \bibinfo{journal}{Phys. Rev. B} \textbf{\bibinfo{volume}{71}},
  \bibinfo{pages}{094505} (\bibinfo{year}{2005}).

\bibitem[{\citenamefont{Bortz and Stolze}(2007)}]{bor06}
\bibinfo{author}{\bibfnamefont{M.}~\bibnamefont{Bortz}} \bibnamefont{and}
  \bibinfo{author}{\bibfnamefont{J.}~\bibnamefont{Stolze}},
  \bibinfo{journal}{J. Stat. Mech.} p. \bibinfo{pages}{P06018}
  (\bibinfo{year}{2007}).

\bibitem[{\citenamefont{Sklyanin}(1999)}]{skl99}
\bibinfo{author}{\bibfnamefont{E.}~\bibnamefont{Sklyanin}},
  \bibinfo{journal}{Lett. Math. Phys.} \textbf{\bibinfo{volume}{47}},
  \bibinfo{pages}{275} (\bibinfo{year}{1999}).

\bibitem[{\citenamefont{Dawson et~al.}(2005)\citenamefont{Dawson, Hines,
  McKenzie, and Milburn}}]{daw05}
\bibinfo{author}{\bibfnamefont{C.}~\bibnamefont{Dawson}},
  \bibinfo{author}{\bibfnamefont{A.}~\bibnamefont{Hines}},
  \bibinfo{author}{\bibfnamefont{R.}~\bibnamefont{McKenzie}}, \bibnamefont{and}
  \bibinfo{author}{\bibfnamefont{G.}~\bibnamefont{Milburn}},
  \bibinfo{journal}{Phys. Rev. A} \textbf{\bibinfo{volume}{71}},
  \bibinfo{pages}{052321} (\bibinfo{year}{2005}).

\bibitem[{\citenamefont{Rossini et~al.}(2006)\citenamefont{Rossini, Calarco,
  Giovannetti, Montangero, and Fazio}}]{ros06}
\bibinfo{author}{\bibfnamefont{D.}~\bibnamefont{Rossini}},
  \bibinfo{author}{\bibfnamefont{T.}~\bibnamefont{Calarco}},
  \bibinfo{author}{\bibfnamefont{V.}~\bibnamefont{Giovannetti}},
  \bibinfo{author}{\bibfnamefont{S.}~\bibnamefont{Montangero}},
  \bibnamefont{and} \bibinfo{author}{\bibfnamefont{R.}~\bibnamefont{Fazio}},
  \bibinfo{journal}{arXiv:quant-ph/0611242}  (\bibinfo{year}{2006}).

\end{thebibliography}
\end{document}